\documentclass{psapm-l}

\newcommand{\ket}[1]{|#1\rangle}      
\newcommand{\bra}[1]{\langle#1|}      
\newcommand{\abs}[1]{|#1|}            
\newcommand{\com}[2]{[#1, #2]}        
\newcommand{\acom}[2]{\{#1, #2\}}     
\newcommand{\set}[1]{\{#1\}}          

\newcommand{\GF}{{\rm GF}(4)}          
\newcommand{\logzero}{\ket{\overline{0}}}  
\newcommand{\logone}{\ket{\overline{1}}}   
\newcommand{\logzerobra}{\bra{\overline{0}}} 
\newcommand{\logonebra}{\bra{\overline{1}}}  
\newcommand{\E}{\mathcal{E}}            
\newcommand{\Hilbert}{\mathcal{H}}          
\newcommand{\Pauli}{\mathcal{P}}            

\newtheorem{thm}{Theorem}

\DeclareMathOperator{\tr}{tr}

\begin{document}

\title{An Introduction to Quantum Error Correction}
\author{Daniel Gottesman}

\address{Microsoft Corporation\\One Microsoft Way\\Redmond, WA
98052}

\email{gottesma@microsoft.com}

\urladdr{http://www.research.microsoft.com/\~{}gottesma}

\subjclass{Primary 81P68; Secondary 94B60}

\begin{abstract}
Quantum states are very delicate, so it is likely some sort of
quantum error correction will be necessary to build reliable
quantum computers. The theory of quantum error-correcting codes
has some close ties to and some striking differences from the
theory of classical error-correcting codes. Many quantum codes can
be described in terms of the stabilizer of the codewords. The
stabilizer is a finite Abelian group, and allows a straightforward
characterization of the error-correcting properties of the code.
The stabilizer formalism for quantum codes also illustrates the
relationships to classical coding theory, particularly classical
codes over $\GF$, the finite field with four elements.
\end{abstract}

\maketitle

\section{Background: the need for error correction}

Quantum computers have a great deal of potential, but to realize
that potential, they need some sort of protection from noise.

Classical computers don't use error correction.  One reason for
this is that classical computers use a large number of electrons,
so when one goes wrong, it is not too serious.  A single qubit in
a quantum computer will probably be just one, or a small number,
of particles, which already creates a need for some sort of error
correction.

Another reason is that classical computers are digital: after each
step, they correct themselves to the closer of 0 or 1.  Quantum
computers have a continuum of states, so it would seem, at first
glance, that they cannot do this. For instance, a likely source of
error is over-rotation: a state $\alpha \ket{0} + \beta \ket{1}$
might be supposed to become $\alpha \ket{0} + \beta e^{i\phi}
\ket{1}$, but instead becomes $\alpha \ket{0} + \beta e^{i(\phi +
\delta)} \ket{1}$.  The actual state is very close to the correct
state, but it is still wrong.  If we don't do something about
this, the small errors will build up over the course of the
computation, and eventually will become a big error.

Furthermore, quantum states are intrinsically delicate: looking at
one collapses it. $\alpha \ket{0} + \beta \ket{1}$ becomes
$\ket{0}$ with probability $\abs{\alpha}^2$ and $\ket{1}$ with
probability $\abs{\beta}^2$.  The environment is constantly trying
to look at the state, a process called {\em decoherence}.  One
goal of quantum error correction will be to prevent the
environment from looking at the data.

There is a well-developed theory of classical error-correcting
codes, but it doesn't apply here, at least not directly.  For one
thing, we need to keep the phase correct as well as correcting bit
flips.  There is another problem, too.  Consider the simplest
classical code, the repetition code:
\begin{align}
  0 & \rightarrow 000
\\1 & \rightarrow 111
\end{align}
It will correct a state such as $010$ to the majority value
(becoming $000$ in this case).\footnote{Actually, a classical
digital computer is using a repetition code -- each bit is encoded
in many electrons (the repetition), and after each time step, it
is returned to the value held by the majority of the electrons
(the error correction).}

We might try a quantum repetition code:
\begin{equation}
\ket{\psi} \rightarrow \ket{\psi} \otimes \ket{\psi} \otimes
\ket{\psi}
\end{equation}
However, no such code exists because of the No-Cloning
theorem~\cite{Dieks,WZ}:

\begin{thm}[No-Cloning]
There is no quantum operation that takes a state $\ket{\psi}$ to
$\ket{\psi} \otimes \ket{\psi}$ for all states $\ket{\psi}$.
\end{thm}

\begin{proof}
This fact is a simple consequence of the linearity of quantum
mechanics.  Suppose we had such an operation and $\ket{\psi}$ and
$\ket{\phi}$ are distinct. Then, by the definition of the
operation,
\begin{align}
\ket{\psi} & \rightarrow \ket{\psi} \ket{\psi}
\\ \ket{\phi} & \rightarrow \ket{\phi} \ket{\phi}
\\ \ket{\psi} + \ket{\phi} & \rightarrow
\left( \ket{\psi} + \ket{\phi} \right) \left( \ket{\psi} +
\ket{\phi} \right). \label{eq:cloned}
\end{align}
(Here, and frequently below, I omit normalization, which is
generally unimportant.)

But by linearity,
\begin{equation}
\ket{\psi} + \ket{\phi} \rightarrow \ket{\psi} \ket{\psi} +
\ket{\phi} \ket{\phi}.
\end{equation}
This differs from~(\ref{eq:cloned}) by the crossterm
\begin{equation}
\ket{\psi} \ket{\phi} + \ket{\phi} \ket{\psi}.
\end{equation}

\end{proof}

\section{The nine-qubit code}

To solve these problems, we will try a variant of the repetition
code~\cite{Shor}.

\begin{align}
\ket{0} & \rightarrow \logzero = \left( \ket{000} + \ket{111}
\right) \left( \ket{000} + \ket{111} \right) \left( \ket{000} +
\ket{111} \right)
\\ \ket{1} & \rightarrow \logone = \left( \ket{000} -
\ket{111} \right) \left( \ket{000} - \ket{111} \right) \left(
\ket{000} - \ket{111} \right)
\end{align}

Note that this does not violate the No-Cloning theorem, since an
arbitrary codeword will be a linear superposition of these two
states
\begin{equation}
\alpha \logzero + \beta \logone \neq \left[ \alpha (\ket{000} +
\ket{111}) + \beta (\ket{000} - \ket{111}) \right]^{\otimes 3}.
\end{equation}
The superposition is linear in $\alpha$ and $\beta$.  The complete
set of codewords for this (or any other) quantum code form a
linear subspace of the Hilbert space, the {\em coding space}.

The inner layer of this code corrects bit flip errors:  We take
the majority within each set of three, so
\begin{equation}
\ket{010} \pm \ket{101} \rightarrow \ket{000} \pm \ket{111}.
\end{equation}
The outer layer corrects phase flip errors:  We take the majority
of the three signs, so
\begin{equation}
(\ket{\cdot} + \ket{\cdot}) (\ket{\cdot} - \ket{\cdot})
(\ket{\cdot} + \ket{\cdot}) \rightarrow (\ket{\cdot} +
\ket{\cdot}) (\ket{\cdot} + \ket{\cdot}) (\ket{\cdot} +
\ket{\cdot}).
\end{equation}
Since these two error correction steps are independent, the code
also works if there is both a bit flip error {\em and} a phase
flip error.

Note that in both cases, we must be careful to measure just what
we want to know and no more, or we would collapse the
superposition used in the code.  I'll discuss this in more detail
in section~\ref{sec:stabilizers}.

The bit flip, phase flip, and combined bit and phase flip errors
are important, so let's take a short digression to discuss them.
We'll also throw in the identity matrix, which is what we get if
no error occurs.  The definitions of these four operators are
given in table~\ref{tab:Pauli}.  The factor of $i$ in the
definition of $Y$ has little practical significance --- overall
phases in quantum mechanics are physically meaningless --- but it
makes some manipulations easier later.  It also makes some
manipulations harder, so either is a potentially reasonable
convention.

\begin{table}
\centering
\begin{tabular}{lll}
Identity & $I = \left( \begin{array}{cc} 1 & 0 \\ 0 & 1
\end{array} \right)$ & $I \ket{a} = \ket{a}$
\\ Bit Flip & $X = \left( \begin{array}{cc} 0 & 1 \\ 1 & 0
\end{array} \right)$ & $X \ket{a} = \ket{a \oplus 1}$
\\ Phase Flip & $Z = \left( \begin{array}{cc} 1 & 0 \\ 0 & -1
\end{array} \right)$ & $Z \ket{a} = (-1)^a \ket{a}$
\\ Bit \& Phase & $Y = \left( \begin{array}{cc} 0 & -i \\ i & 0
\end{array} \right) = iXZ$ & $Y \ket{a} = i (-1)^a \ket{a \oplus 1}$
\end{tabular}
\caption{The Pauli matrices} \label{tab:Pauli}
\end{table}

The group generated by tensor products of these $4$ operators is
called the Pauli group.  $X$, $Y$, and $Z$ anticommute: $XZ = -ZX$
(also written $\acom{X}{Z} = 0$).  Similarly, $\acom{X}{Y} = 0$
and $\acom{Y}{Z} = 0$.  Thus, the $n$-qubit Pauli group $\Pauli_n$
consists of the $4^n$ tensor products of $I$, $X$, $Y$, and $Z$,
and an overall phase of $\pm 1$ or $\pm i$, for a total of
$4^{n+1}$ elements.  The phase of the operators used is not
generally very important, but we can't discard it completely.  For
one thing, the fact that this is not an Abelian group is quite
important, and we would lose that if we dropped the phase!

$\Pauli_n$ is useful because of its nice algebraic properties. Any
pair of elements of $\Pauli_n$ either commute or anticommute.
Also, the square of any element of $\Pauli_n$ is $\pm 1$.  We
shall only need to work with the elements with square $+1$, which
are tensor products of $I$, $X$, $Y$, and $Z$ with an overall sign
$\pm 1$; the phase $i$ is only necessary to make $\Pauli_n$ a
group.  Define the {\em weight} of an operator in $\Pauli_n$ to be
the number of tensor factors which are not $I$.  Thus, $X \otimes
Y \otimes I$ has weight $2$.

Another reason the Pauli matrices are important is that they span
the space of $2 \times 2$ matrices, and the $n$-qubit Pauli group
spans the space of $2^n \times 2^n$ matrices.  For instance, if we
have a general phase error
\begin{equation}
R_{\theta/2} = \left( \begin{array}{cc} 1 & 0 \\ 0 & e^{i\theta}
\end{array} \right) = e^{i \theta/2} \left( \begin{array}{cc}
e^{-i \theta/2} & 0 \\ 0 & e^{i \theta/2} \end{array} \right)
\end{equation}
(again, the overall phase does not matter), we can write it as
\begin{equation}
R_{\theta/2} = \cos \frac{\theta}{2} \ I - i \sin \frac{\theta}{2}
\ Z.
\end{equation}

It turns out that our earlier error correction procedure will also
correct this error, without any additional effort.  For instance,
the earlier procedure might use some extra qubits ({\em ancilla}
qubits) that are initialized to $\ket{0}$ and record what type of
error occurred. Then we look at the ancilla and invert the error
it tells us:
\begin{align}
Z \left( \alpha \logzero + \beta \logone \right) \otimes
\ket{0}_{\rm anc} & \rightarrow Z \left( \alpha \logzero + \beta
\logone \right) \otimes \ket{Z}_{\rm anc}
\\ & \rightarrow \left( \alpha \logzero + \beta
\logone \right) \otimes \ket{Z}_{\rm anc}
\\ I \left( \alpha \logzero + \beta \logone \right) \otimes
\ket{0}_{\rm anc} & \rightarrow I \left( \alpha \logzero + \beta
\logone \right) \otimes \ket{\text{no error}}_{\rm anc}
\\ & \rightarrow \left( \alpha \logzero + \beta
\logone \right) \otimes \ket{\text{no error}}_{\rm anc}
\end{align}

When the actual error is $R_{\theta/2}$, recording the error in
the ancilla gives us a superposition:
\begin{equation}
\cos \frac{\theta}{2} \ I \left( \alpha \logzero + \beta \logone
\right) \otimes \ket{\text{no error}}_{\rm anc} - i \sin
\frac{\theta}{2} \ Z \left( \alpha \logzero + \beta \logone
\right) \otimes \ket{Z}_{\rm anc}
\end{equation}
Then we measure the ancilla, which with probability $\sin^2
\theta/2$ gives us
\begin{equation}
Z \left( \alpha \logzero + \beta \logone \right) \otimes
\ket{Z}_{\rm anc},
\end{equation}
and with probability $\cos^2 \theta/2$ gives us
\begin{equation}
I \left( \alpha \logzero + \beta \logone \right) \otimes
\ket{\text{no error}}_{\rm anc}.
\end{equation}
In each case, inverting the error indicated in the ancilla
restores the original state.

It is easy to see this argument works for any linear combination
of errors~\cite{Shor,Steane:7}:

\begin{thm}
\label{thm:linear} If a quantum code corrects errors $A$ and $B$,
it also corrects any linear combination of $A$ and $B$.  In
particular, if it corrects all weight $t$ Pauli errors, then the
code corrects all $t$-qubit errors.
\end{thm}

So far, we have only considered individual unitary errors that
occur on the code.  But we can easily add in all possible quantum
errors.  The most general quantum operation, including
decoherence, interacts the quantum state with some extra qubits
via a unitary operation, then discards some qubits.  This process
can turn pure quantum states into mixed quantum states, which are
normally described using density matrices.  We can write the most
general operation as a transformation on density matrices
\begin{equation}
\rho \rightarrow \sum_i E_i \rho E_i^\dagger,
\end{equation}
where the $E_i$s are normalized so $\sum E_i^\dagger E_i = I$. The
density matrix $\rho$ can be considered to represent an ensemble
of pure quantum states $\ket{\psi}$, each of which, in this case,
should be in the coding space of the code.  Then this operation
simply performs the following operation on each $\ket{\psi}$:
\begin{equation}
\ket{\psi} \rightarrow E_i \ket{\psi} \text{ with probability }
\abs{E_i \ket{\psi}}^2.
\end{equation}
If we can correct each of the individual errors $E_i$, then we can
correct this general error as well.  For instance, for quantum
operations that only affect a single qubit of the code, $E_i$ will
necessarily be in the linear span of $I$, $X$, $Y$, and $Z$, so we
can correct it.  Thus, in the statement of
theorem~\ref{thm:linear}, ``all $t$-qubit errors'' really does
apply to {\em all} $t$-qubit errors, not just unitary ones.

We can go even further.  It is not unreasonable to expect that
every qubit in our nine-qubit code will be undergoing some small
error. For instance, qubit $i$ experiences the error $I + \epsilon
E_i$, where $E_i$ is some single-qubit error.  Then the overall
error is
\begin{equation}
\bigotimes (I + \epsilon E_i) = I + \epsilon \left(E_1 \otimes
I^{\otimes 8} + I \otimes E_2 \otimes I^{\otimes 7} + \ldots
\right) + O(\epsilon^2)
\end{equation}

That is, to order $\epsilon$, the actual error is the sum of
single-qubit errors, which we know the nine-qubit code can
correct.  That means that after the error correction procedure,
the state will be correct to $O(\epsilon^2)$ (when the two-qubit
error terms begin to become important).  While the code cannot
completely correct this error, it still produces a significant
improvement over not doing error correction when $\epsilon$ is
small.  A code correcting more errors would do even better.

\section{General properties of quantum error-correcting codes}

Let us try to understand what properties are essential to the
success of the nine-qubit code, and derive conditions for a
subspace to form a quantum error-correcting code.

One useful feature was {\em linearity}, which will be true of any
quantum code. We only need to correct a basis of errors ($I$, $X$,
$Y$, and $Z$ in the one-qubit case), and all other errors will
follow, as per theorem~\ref{thm:linear}.

In any code, we must never confuse $\logzero$ with $\logone$, even
in the presence of errors.  That is, $E \logzero$ is orthogonal to
$F \logone$:
\begin{equation}
\logzerobra E^\dagger F \logone = 0. \label{eq:orthog}
\end{equation}

It is {\em sufficient} to distinguish error $E$ from error $F$
when they act on $\logzero$ and $\logone$.  Then a measurement
will tell us exactly what the error is and we can correct it:
\begin{equation}
\logzerobra E^\dagger F \logzero = \logonebra E^\dagger F \logone
= 0 \label{eq:nondeg}
\end{equation}
for $E \neq F$.

But (\ref{eq:nondeg}) is not {\em necessary}: in the nine-qubit
code, we cannot distinguish between $Z_1$ and $Z_2$, but that is
OK, since we can correct either one with a single operation.  To
understand the necessary condition, it is helpful to look at the
operators $F_1 = (Z_1 + Z_2)/2$ and $F_2 = (Z_1 - Z_2)/2$ instead
of $Z_1$ and $Z_2$.  $F_1$ and $F_2$ span the same space as $Z_1$
and $Z_2$, so Shor's code certainly corrects them; let us try to
understand how. When we use the $F$s as the basis errors, now
equation~(\ref{eq:nondeg}) {\em is} satisfied.  That means we can
make a measurement and learn what the error is.  We also have to
invert it, and this is a potential problem, since $F_1$ and $F_2$
are not unitary.  However, $F_1$ acts the same way as $Z_1$ on the
coding space, so $Z_1^\dagger$ suffices to invert $F_1$ on the
states of interest.  $F_2$ acts the same way as the $0$ operator
on the coding space.  We can't invert this, but we don't need to
--- since $F_2$ annihilates codewords, it can never contribute a
component to the actual state of the system.

The requirement to invert the errors produces a third condition:
\begin{equation}
\logzerobra E^\dagger E \logzero = \logonebra E^\dagger E \logone.
\label{eq:equiv}
\end{equation}
Either this value is nonzero, as for $F_1$, in which case some
unitary operator will act the same way as $E$ on the coding space,
or it will be zero, as for $F_2$, in which case $E$ annihilates
codewords and never arises.

These arguments show that if there is some basis for the space of
errors for which equations (\ref{eq:orthog}), (\ref{eq:nondeg}),
and (\ref{eq:equiv}) hold, then the states $\logzero$ and
$\logone$ span a quantum error-correcting code.  Massaging these
three equations together and generalizing to multiple encoded
qubits, we get the following theorem~\cite{BDSW,KL}:

\begin{thm}
Suppose $\E$ is a linear space of errors acting on the Hilbert
space $\Hilbert$. Then a subspace $C$ of $\Hilbert$ forms a
quantum error-correcting code correcting the errors $\E$ iff
\begin{equation}
\bra{\psi} E^\dagger E \ket{\psi} = C(E) \label{eq:QECC}
\end{equation}
for all $E \in \E$.  The function $C(E)$ does not depend on the
state $\ket{\psi}$.
\end{thm}

\begin{proof}

Suppose $\set{E_a}$ is a basis for $\E$ and $\set{\ket{\psi_i}}$
is a basis for $C$.  By setting $E$ and $\ket{\psi}$ equal to the
basis elements and to the sum and difference of two basis elements
(with or without a phase factor $i$), we can see that
(\ref{eq:QECC}) is equivalent to
\begin{equation}
\bra{\psi_i} E_a^\dagger E_b \ket{\psi_j} = C_{ab} \delta_{ij},
\label{eq:QECC2}
\end{equation}
where $C_{ab}$ is a Hermitian matrix independent of $i$ and $j$.

Suppose equation~(\ref{eq:QECC2}) holds.  We can diagonalize
$C_{ab}$. This involves choosing a new basis $\set{F_a}$ for $\E$,
and the result is equations (\ref{eq:orthog}), (\ref{eq:nondeg}),
and (\ref{eq:equiv}).  The arguments before the theorem show that
we can measure the error, determine it uniquely (in the new
basis), and invert it (on the coding space). Thus, we have a
quantum error-correcting code.

Now suppose we have a quantum error-correcting code, and let
$\ket{\psi}$ and $\ket{\phi}$ be two distinct codewords.  Then we
must have
\begin{equation}
\bra{\psi} E^\dagger E \ket{\psi} = \bra{\phi} E^\dagger E
\ket{\phi}
\end{equation}
for all $E$.  That is, (\ref{eq:QECC}) must hold. If not, $E$
changes the relative size of $\ket{\psi}$ and $\ket{\phi}$.  Both
$\ket{\psi} + \ket{\phi}$ and $\ket{\psi} + c \ket{\phi}$ are
valid codewords, and
\begin{equation}
E (\ket{\psi} + \ket{\phi}) = N (\ket{\psi} + c \ket{\phi}),
\end{equation}
where $N$ is a normalization factor and
\begin{equation}
c = \bra{\psi} E^\dagger E \ket{\psi} / \bra{\phi} E^\dagger E
\ket{\phi}.
\end{equation}
The error $E$ will actually change the encoded state, which is a
failure of the code, unless $c=1$.

\end{proof}

There is a slight subtlety to the phrasing of
equation~(\ref{eq:QECC}).  We require $\E$ to be a linear space of
errors, which means that it must be closed under sums of errors
which may act on different qubits.  In contrast, for a code that
corrects $t$ errors, in~(\ref{eq:QECC2}), it is safe to consider
only $E_a$ and $E_b$ acting on just $t$ qubits.  We can restrict
even further, and only use Pauli operators as $E_a$ and $E_b$,
since they will span the space of $t$-qubit errors.  This leads us
to a third variation of the condition:
\begin{equation}
\bra{\psi} E \ket{\psi} = C'(E), \label{eq:QECC3}
\end{equation}
where $E$ is now any operator acting on $2t$ qubits (that is, it
replaces $E_a^\dagger E_b$ in (\ref{eq:QECC2})).  This can be
easily interpreted as saying that no measurement on $2t$ qubits
can learn information about the codeword.  Alternatively, it says
we can {\em detect} up to $2t$ errors on the code without
necessarily being able to say what those errors are.  That is, we
can distinguish those errors from the identity.

If the matrix $C_{ab}$ in (\ref{eq:QECC2}) has maximum rank, the
code is called {\em nondegenerate}.  If not, as for the nine-qubit
code, the code is {\em degenerate}.  In a degenerate code,
different errors look the same when acting on the coding subspace.

For a nondegenerate code, we can set a simple bound on the
parameters of the code simply by counting states.  Each error $E$
acting on each basis codeword $\ket{\psi_i}$ produces a linearly
independent state.  All of these states must fit in the full
Hilbert space of $n$ qubits, which has dimension $2^n$.  If the
code encodes $k$ qubits, and corrects errors on up to $t$ qubits,
then
\begin{equation}
\left( \sum_{j=0}^t 3^j \binom{n}{j} \right) 2^k \leq 2^n.
\label{eq:QHB}
\end{equation}
The quantity in parentheses is the number of errors of {\em
weight} $t$ or less: that is, the number of tensor products of
$I$, $X$, $Y$, and $Z$ that are the identity in all but $t$ or
fewer places.  This inequality is called the {\em quantum Hamming
bound}.  While the quantum Hamming bound only applies to
nondegenerate codes, we do not know of any codes that beat it.

For $t=1$, $k=1$, the quantum Hamming bound tells us $n \geq 5$.
In fact, there is a code with $n=5$, which you will see later.  A
code that corrects $t$ errors is said to have {\em distance}
$2t+1$, because it takes $2t+1$ single-qubit changes to get from
one codeword to another.  We can also define distance as the
minimum weight of an operator $E$ that violates equation
(\ref{eq:QECC3}) (a definition which also allows codes of even
distance). A quantum code using $n$ qubits to encode $k$ qubits
with distance $d$ is written as an $[[n, k, d]]$ code (the double
brackets distinguish it from a classical code). Thus, the
nine-qubit code is a $[[9, 1, 3]]$ code, and the five-qubit code
is a $[[5, 1, 3]]$ code.

We can also set a lower bound telling us when codes exist.  I will
not prove this here, but an $[[n, k, d]]$ code exists when
\begin{equation}
\left( \sum_{j=0}^{d-1} 3^j \binom{n}{j} \right) 2^k \leq 2^n
\label{eq:QGV}
\end{equation}
(known as the quantum Gilbert-Varshamov bound~\cite{CRSS}).  This
differs from the quantum Hamming bound in that the sum goes up to
$d-1$ (which is equal to $2t$) rather than stopping at $t$.

\begin{thm}
A quantum $[[n, k, d]]$ code exists when (\ref{eq:QGV})~holds. Any
nondegenerate $[[n, k, d]]$ code must satisfy~(\ref{eq:QHB}).  For
large $n$, $R = k/n$ and $p = d/2n$ fixed, the best nondegenerate
quantum codes satisfy
\begin{equation}
1 - 2p \log_2 3 - H(2p) \leq R \leq 1 - p \log_2 3 - H(p),
\end{equation}
where $H(x) = - x \log_2 x - (1-x) \log_2 (1-x)$.
\end{thm}

One further bound, known as the Knill-Laflamme bound~\cite{KL} or
the quantum Singleton bound, applies even to degenerate quantum
codes. For an $[[n,k,d]]$ quantum code,
\begin{equation}
n - k \geq 2d - 2.
\end{equation}
This shows that the $[[5,1,3]]$ code really is optimal --- a
$[[4,1,3]]$ code would violate this bound.

I will not prove the general case of this bound, but the case of
$k=1$ can be easily understood as a consequence of the No-Cloning
theorem.  Suppose $r$ qubits of the code are missing.  We can
substitute $\ket{0}$ states for the missing qubits, but there are
$r$ errors on the resulting codeword.  The errors are of unknown
type, but all the possibilities are on the same set of $r$ qubits.
Thus, all products $E_a^\dagger E_b$ in condition~(\ref{eq:QECC2})
have weight $r$ or less, so this sort of error (an ``erasure''
error~\cite{GBP}) can be corrected by a code of distance $r+1$.
Now suppose we had an $[[n, 1, d]]$ code with $n \leq 2d-2$.  Then
we could split the qubits in the code into two groups of size at
most $d-1$. Each group would have been subject to at most $d-1$
erasure errors, and could therefore be corrected without access to
the other group. This would produce two copies of the encoded
state, which we know is impossible.

\section{Stabilizer codes}
\label{sec:stabilizers}

Now let us return to the nine-qubit code, and examine precisely
what we need to do to correct errors.

First, we must determine if the first three qubits are all the
same, and if not, which is different.  We can do this by measuring
the parity of the first two qubits and the parity of the second
and third qubits.  That is, we measure
\begin{equation}
Z \otimes Z \otimes I \text{ and } I \otimes Z \otimes Z.
\end{equation}
The first tells us if an $X$ error has occurred on qubits one or
two, and the second tells us if an $X$ error has occurred on
qubits two or three.  Note that the error detected in both cases
anticommutes with the error measured.  Combining the two pieces of
information tells us precisely where the error is.

We do the same thing for the other two sets of three.  That gives
us four more operators to measure.  Note that measuring $Z \otimes
Z$ gives us just the information we want and no more.  This is
crucial so that we do not collapse the superpositions used in the
code.  We can do this by bringing in an ancilla qubit.  We start
it in the state $\ket{0} + \ket{1}$ and perform controlled-$Z$
operations to the first and second qubits of the code:
\begin{align}
\left( \ket{0} + \ket{1} \right) \sum_{abc} c_{abc} \ket{abc} &
\rightarrow \sum_{abc} c_{abc} \left( \ket{0} \ket{abc} + (-1)^{a
\oplus b} \ket{1} \ket{abc} \right)
\\ & = \sum_{abc} c_{abc} \left( \ket{0} + (-1)^{{\rm parity} (a,b)}
\ket{1} \right) \ket{abc}.
\end{align}
At this point, measuring the ancilla in the basis $\ket{0} \pm
\ket{1}$ will tell us the eigenvalue of $Z \otimes Z \otimes I$,
but nothing else about the data.

Second, we must check if the three signs are the same or
different.  We do this by measuring
\begin{equation}
X \otimes X \otimes X \otimes X \otimes X \otimes X \otimes I
\otimes I \otimes I
\end{equation}
and
\begin{equation}
I \otimes I \otimes I \otimes X \otimes X \otimes X \otimes X
\otimes X \otimes X.
\end{equation}
This gives us a total of $8$ operators to measure.  These two
measurements detect $Z$ errors on the first six and last six
qubits, correspondingly.  Again note that the error detected
anticommutes with the operator measured.

This is no coincidence: in each case, we are measuring an operator
$M$ which should have eigenvalue $+1$ for any codeword:
\begin{equation}
M \ket{\psi} = \ket{\psi}.
\end{equation}
If an error $E$ which anticommutes with $M$ has occurred, then the
true state is $E \ket{\psi}$, and
\begin{equation}
M \left( E \ket{\psi} \right) = - E M \ket{\psi} = - E \ket{\psi}.
\end{equation}
That is, the new state has eigenvalue $-1$ instead of $+1$.  We
use this fact to correct errors: each single-qubit error $E$
anticommutes with a particular set of operators $\set{M}$; which
set, exactly, tells us what $E$ is.

In the case of the nine-qubit code, we cannot tell exactly what
$E$ is, but it does not matter.  For instance, we cannot
distinguish $Z_1$ and $Z_2$ because
\begin{equation}
Z_1 Z_2 \ket{\psi} = \ket{\psi} \ \ \Longleftrightarrow \ \ Z_1
\ket{\psi} = Z_2 \ket{\psi}.
\end{equation}
This is an example of the fact that the nine-qubit code is
degenerate.

Table~\ref{tab:nine} summarizes the operators we measured.
\begin{table}
\centering
\begin{tabular}{ccccccccc}
   $Z$ & $Z$ & $I$ & $I$ & $I$ & $I$ & $I$ & $I$ & $I$
\\ $I$ & $Z$ & $Z$ & $I$ & $I$ & $I$ & $I$ & $I$ & $I$
\\ $I$ & $I$ & $I$ & $Z$ & $Z$ & $I$ & $I$ & $I$ & $I$
\\ $I$ & $I$ & $I$ & $I$ & $Z$ & $Z$ & $I$ & $I$ & $I$
\\ $I$ & $I$ & $I$ & $I$ & $I$ & $I$ & $Z$ & $Z$ & $I$
\\ $I$ & $I$ & $I$ & $I$ & $I$ & $I$ & $I$ & $Z$ & $Z$
\\ $X$ & $X$ & $X$ & $X$ & $X$ & $X$ & $I$ & $I$ & $I$
\\ $I$ & $I$ & $I$ & $X$ & $X$ & $X$ & $X$ & $X$ & $X$
\end{tabular}
\caption{The stabilizer for the nine-qubit code.  Each column
represents a different qubit.} \label{tab:nine}
\end{table}
These $8$ operators generate an Abelian group called the {\em
stabilizer} of the nine-qubit code.  The stabilizer contains all
operators $M$ in the Pauli group for which $M \ket{\psi} =
\ket{\psi}$ for all $\ket{\psi}$ in the code.

Conversely, given an Abelian subgroup $S$ of the Pauli group
$\Pauli_n$ (which, if you recall, consists of tensor products of
$I$, $X$, $Y$, and $Z$ with an overall phase of $\pm 1, \pm i$),
we can define a quantum code $T(S)$ as the set of states
$\ket{\psi}$ for which $M \ket{\psi} = \ket{\psi}$ for all $M \in
S$.  $S$ must be Abelian and cannot contain $-1$, or the code is
trivial: If $M, N \in S$,
\begin{align}
MN \ket{\psi} = M \ket{\psi} & = \ket{\psi}
\\ NM \ket{\psi} = N \ket{\psi} & = \ket{\psi}
\end{align}
so
\begin{equation}
\com{M}{N} \ket{\psi} = MN \ket{\psi} - NM \ket{\psi} = 0.
\end{equation}
Since elements of the Pauli group either commute or anticommute,
$\com{M}{N} = 0$.  Clearly, if $M = -1 \in S$, there is no
nontrivial $\ket{\psi}$ for which $M \ket{\psi} = \ket{\psi}$.

If these conditions are satisfied, there will be a nontrivial
subspace consisting of states fixed by all elements of the
stabilizer.  We can tell how many errors the code corrects by
looking at operators that commute with the stabilizer. We can
correct errors $E$ and $F$ if either $E^\dagger F \in S$ (so $E$
and $F$ act the same on codewords), or if $\exists M \in S\ {\rm
s.t.}\ \acom{M}{E^\dagger F} = 0$, in which case measuring the
operator $M$ distinguishes between $E$ and $F$. If the first
condition is ever true, the stabilizer code is degenerate;
otherwise it is nondegenerate.

We can codify this by looking at the normalizer $N(S)$ of $S$ in
the Pauli group (which is in this case equal to the centralizer,
composed of Pauli operators which commute with $S$).  The distance
$d$ of the code is the minimum weight of any operator in $N(S)
\setminus S$~\cite{CRSS,Gottesman}.

\begin{thm}
Let $S$ be an Abelian subgroup of order $2^a$ of the $n$-qubit
Pauli group, and suppose $-1 \not\in S$.  Let $d$ be the minimum
weight of an operator in $N(S) \setminus S$.  Then the space of
states $T(S)$ stabilized by all elements of $S$ is an $[[n, n-a,
d]]$ quantum code.
\end{thm}

To correct errors of weight $(d-1)/2$ or below, we simply measure
the generators of $S$.  This will give us a list of eigenvalues,
the {\em error syndrome}, which tells us whether the error $E$
commutes or anticommutes with each of the generators.  The error
syndromes of $E$ and $F$ are equal iff the error syndrome of
$E^\dagger F$ is trivial.  For a nondegenerate code, the error
syndrome uniquely determines the error $E$ (up to a trivial
overall phase) --- the generator that anticommutes with $E^\dagger
F$ distinguishes $E$ from $F$. For a degenerate code, the error
syndrome is not unique, but error syndromes are only repeated when
$E^\dagger F \in S$, implying $E$ and $F$ act the same way on the
codewords.

If the stabilizer has $a$ generators, then the code encodes $n-a$
qubits.  Each generator divides the allowed Hilbert space into
$+1$ and $-1$ eigenspaces of equal sizes. To prove the statement,
note that we can find an element $G$ of the Pauli group that has
any given error syndrome (though $G$ may have weight greater than
$(d-1)/2$, or even greater than $d$). Each $G$ maps $T(S)$ into an
orthogonal but isomorphic subspace, and there are $2^a$ possible
error syndromes, so $T(S)$ has dimension at most $2^n/2^a$.  In
addition, the Pauli group spans $U(2^n)$, so its orbit acting on
any single state contains a basis for $\Hilbert$. Every Pauli
operator has {\em some} error syndrome, so $T(S)$ has dimension
exactly $2^{n-a}$.

\section{Some other important codes}

Stabilizers make it easy to describe new codes.  For instance, we
can start from classical coding theory, which describes a linear
code by a generator matrix or its dual, the parity check matrix.
Each row of the generator matrix is a codeword, and the other
codewords are all linear combinations of the rows of the generator
matrix.  The rows of the parity check matrix specify parity checks
all the classical codewords must satisfy.  (In quantum codes, the
stabilizer is closely analogous to the classical parity check
matrix.)  One well-known code is the seven-bit Hamming code
correcting one error, with parity check matrix
\begin{equation}
\left( \begin{array}{ccccccc}
   1 & 1 & 1 & 1 & 0 & 0 & 0
\\ 1 & 1 & 0 & 0 & 1 & 1 & 0
\\ 1 & 0 & 1 & 0 & 1 & 0 & 1
\end{array} \right).
\end{equation}

If we replace each $1$ in this matrix by the operator $Z$, and $0$
by $I$, we are really changing nothing, just specifying three
operators that implement the parity check measurements.  The
statement that the classical Hamming code corrects one error is
the statement that each bit flip error of weight one or two
anticommutes with one of these three operators.

Now suppose we replace each $1$ by $X$ instead of $Z$.  We again
get three operators, and they will anticommute with any weight one
or two $Z$ error.  Thus, if we make a stabilizer out of the three
$Z$ operators and the three $X$ operators, as in
table~\ref{tab:seven}, we get a code that can correct any single
qubit error~\cite{Steane:7}.  $X$ errors are picked up by the
first three generators, $Z$ errors by the last three, and $Y$
errors are distinguished by showing up in both halves.  Of course,
there is one thing to check: the stabilizer must be Abelian; but
that is easily verified.
\begin{table}
\centering
\begin{tabular}{ccccccc}
   $Z$ & $Z$ & $Z$ & $Z$ & $I$ & $I$ & $I$
\\ $Z$ & $Z$ & $I$ & $I$ & $Z$ & $Z$ & $I$
\\ $Z$ & $I$ & $Z$ & $I$ & $Z$ & $I$ & $Z$
\\ $X$ & $X$ & $X$ & $X$ & $I$ & $I$ & $I$
\\ $X$ & $X$ & $I$ & $I$ & $X$ & $X$ & $I$
\\ $X$ & $I$ & $X$ & $I$ & $X$ & $I$ & $X$
\end{tabular}
\caption{Stabilizer for the seven-qubit code.} \label{tab:seven}
\end{table}
The stabilizer has $6$ generators on $7$ qubits, so it encodes $1$
qubit --- it is a $[[7, 1, 3]]$ code.

In this example, we used the same classical code for both the $X$
and $Z$ generators, but there was no reason we had to do so.  We
could have used any two classical codes $C_1$ and
$C_2$~\cite{CS,Steane}. The only requirement is that the $X$ and
$Z$ generators commute. This corresponds to the statement that
$C_2^\perp \subseteq C_1$ ($C_2^\perp$ is the dual code to $C_2$,
consisting of those words which are orthogonal to the codewords of
$C_2$).  If $C_1$ is an $[n, k_1, d_1]$ code, and $C_2$ is an $[n,
k_2, d_2]$ code (recall single brackets means a classical code),
then the corresponding quantum code is an $[[n, k_1 + k_2 - n,
\min(d_1, d_2)]]$ code.\footnote{In fact, the true distance of the
code could be larger than expected because of the possibility of
degeneracy, which would not have been a factor for the classical
codes.} This construction is known as the CSS construction after
its inventors Calderbank, Shor, and Steane.

The codewords of a CSS code have a particularly nice form.  They
all must satisfy the same parity checks as the classical code
$C_1$, so all codewords will be superpositions of words of $C_1$.
The parity check matrix of $C_2$ is the generator matrix of
$C_2^\perp$, so the $X$ generators of the stabilizer add a word of
$C_2^\perp$ to the state.  Thus, the codewords of a CSS code are
of the form
\begin{equation}
\sum_{w \in C_2^\perp} \ket{u + w},
\end{equation}
where $u \in C_1$ ($C_2^\perp \subseteq C_1$, so $u + w \in C_1$).
If we perform a Hadamard transform
\begin{align}
   \ket{0} & \longleftrightarrow \ket{0} + \ket{1}
\\ \ket{1} & \longleftrightarrow \ket{0} - \ket{1}
\end{align}
on each qubit of the code, we switch the $Z$ basis with the $X$
basis, and $C_1$ with $C_2$, so the codewords are now
\begin{equation}
\sum_{w \in C_1^\perp} \ket{u + w} \quad (u \in C_2).
\end{equation}
Thus, to correct errors for a CSS code, we can measure the
parities of $C_1$ in the $Z$ basis, and the parities of $C_2$ in
the $X$ basis.

Another even smaller quantum code is the $[[5, 1, 3]]$ code I
promised earlier~\cite{BDSW,LMPZ}.  Its stabilizer is given in
table~\ref{tab:five}.
\begin{table}
\centering
\begin{tabular}{ccccc}
   $X$ & $Z$ & $Z$ & $X$ & $I$
\\ $I$ & $X$ & $Z$ & $Z$ & $X$
\\ $X$ & $I$ & $X$ & $Z$ & $Z$
\\ $Z$ & $X$ & $I$ & $X$ & $Z$
\end{tabular}
\caption{The stabilizer for the five-qubit code.} \label{tab:five}
\end{table}
I leave it to you to verify that it commutes and actually does
have distance $3$.  You can also work out the codewords.  Since
multiplication by $M \in S$ merely rearranges elements of the
group $S$, the sum
\begin{equation}
\left( \sum_{M \in S} M \right) \ket{\phi} \label{eq:proj}
\end{equation}
is in the code for any state $\ket{\phi}$.  You only need find two
states $\ket{\phi}$ for which (\ref{eq:proj})~is nonzero.  Note
that as well as telling us about the error-correcting properties
of the code, the stabilizer provides a more compact notation for
the coding subspace than listing the basis codewords.

A representation of stabilizers that is often useful is as a pair
of binary matrices, frequently written adjacent with a line
between them~\cite{CRSS}.  The first matrix has a $1$ everywhere
the stabilizer has an $X$ or a $Y$, and a $0$ elsewhere; the
second matrix has a $1$ where the stabilizer has a $Y$ or a $Z$.
Multiplying together Pauli operators corresponds to adding the two
rows for both matrices.  Two operators $M$ and $N$ commute iff
their binary vector representations $(a_1 | b_1)$, $(a_2, b_2)$
are orthogonal under a symplectic inner product: $a_1 b_2 + b_1
a_2 = 0$.  For instance, the stabilizer for the five-qubit code
becomes the matrix
\begin{equation}
\left(
\begin{array}{ccccc}
  $1$ & $0$ & $0$ & $1$ & $0$
\\$0$ & $1$ & $0$ & $0$ & $1$
\\$1$ & $0$ & $1$ & $0$ & $0$
\\$0$ & $1$ & $0$ & $1$ & $0$
\end{array}
\right| \left.
\begin{array}{ccccc}
  $0$ & $1$ & $1$ & $0$ & $0$
\\$0$ & $0$ & $1$ & $1$ & $0$
\\$0$ & $0$ & $0$ & $1$ & $1$
\\$1$ & $0$ & $0$ & $0$ & $1$
\end{array}
\right).
\end{equation}

\section{Codes over $\GF$}

I will finish by describing another connection to classical coding
theory.  Frequently, classical coding theorists consider not just
binary codes, but codes over larger finite fields.  One of the
simplest is $\GF$, the finite field with four elements.  It is a
field of characteristic $2$, containing the elements $\{0, 1,
\omega, \omega^2\}$.
\begin{equation}
\omega^3 = 1,\ \omega + \omega^2 = 1
\end{equation}
It is also useful to consider two operations on $\GF$.  One is
conjugation, which switches the two roots of the characteristic
polynomial $x^2 + x + 1$:
\begin{align}
\overline{1} & = 1  & \overline{\omega} & = \omega^2
\\ \overline{0} & = 0  & \overline{\omega^2} & = \omega
\end{align}
The other is trace.  $\tr x$ is the trace of the linear operator
``multiplication by $x$'' when $\GF$ is considered as a vector
space over $\mathbb{Z}_2$:
\begin{align}
\tr 0 = \tr 1 & = 0
\\ \tr \omega = \tr \omega^2 & = 1
\end{align}

Stabilizer codes make extensive use of the Pauli group $\Pauli_n$.
We can make a connection between stabilizer codes and codes over
$\GF$ by identifying the four operators $I$, $X$, $Y$, and $Z$
with the four elements of $\GF$, as in
table~\ref{tab:GF4}~\cite{CRSS:GF4}.
\begin{table}
\centering
\begin{tabular}{ll}
Stabilizers & $\GF$
\\ \hline
\\ $I$ & $0$
\\ $Z$ & $1$
\\ $X$ & $\omega$
\\ $Y$ & $\omega^2$
\\ tensor products & vectors
\\
\\ multiplication & addition
\\ $\com{M}{N} = 0$ & $\tr (M \cdot \overline{N}) = 0$
\\ $N(S)$ & dual
\end{tabular}
\caption{Connections between stabilizer codes and codes over
$\GF$.} \label{tab:GF4}
\end{table}

The commutativity constraint in the Pauli group becomes a
symplectic inner product between vectors in $\GF$.  The fact that
the stabilizer is Abelian can be phrased in the language of $\GF$
as the fact that the code must be contained in its dual with
respect to this inner product.  To determine the number of errors
corrected by the code, we must examine vectors which are in the
dual (corresponding to $N(S)$) but not in the code (corresponding
to $S$).

The advantage of making this correspondence is that a great deal
of classical coding theory instantly becomes available.  Many
classical codes over $\GF$ are known, and many of them are
self-dual with respect to the symplectic inner product, so they
define quantum codes.  For instance, the five-qubit code is one
such --- in fact, it is just a Hamming code over $\GF$!  Of
course, mostly classical coding theorists consider {\em linear}
codes (which are closed under addition and scalar multiplication),
whereas in the quantum case we wish to consider the slightly more
general class of {\em additive} $\GF$ codes (that is, codes which
are closed under addition of elements, but not necessarily scalar
multiplication).

\section{Fault-Tolerant Quantum Computation}

Hopefully, this paper has given you an understanding of quantum
error-correct\-ing codes, but there is still a major hurdle before
the goal of making quantum computers resistant to errors. You must
also understand how to perform operations on a state encoded in a
quantum code without losing the code's protection against errors,
and how to safely perform error correction when the gates used are
themselves noisy.  For a full discussion of this problem and its
resolutions, see~\cite{Preskill:FT1} or \cite{Preskill:FT2}.

Shor presented the first protocols for fault-tolerant quantum
computation~\cite{Shor:FT}.  While those protocols can be extended
to work for arbitrary stabilizer codes, including those with
multiple encoded qubits per block~\cite{Gottesman:FT}, the gates
which can be performed easily on the code arise from symmetries of
the stabilizer.  The stabilizer of the seven-qubit code has a
particularly large symmetry group and therefore is particularly
good for fault-tolerant computation.

When the error rate per gate is low enough, encoding a state in a
quantum code and performing fault-tolerant operations will reduce
the effective error rate. By concatenating the seven-qubit code or
another code (i.e., encoding each qubit of the code with another
copy of the seven-qubit code, and possibly repeating the procedure
multiple times), we can compound this improvement, giving a
threshold
result~\cite{AB:Threshold,Kitaev:Threshold,KLZ:Threshold}: if the
error rate is below some threshold value, concatenating a code
allows us to perform arbitrarily long fault-tolerant quantum
computations, with overhead that is polylogarithmic in the length
of the computation.

\section{Summary (Quantum Error Correction Sonnet)}

\begin{verse}
We cannot clone, perforce; instead, we split
\\Coherence to protect it from that wrong
\\That would destroy our valued quantum bit
\\And make our computation take too long.

Correct a flip and phase - that will suffice.
\\If in our code another error's bred,
\\We simply measure it, then God plays dice,
\\Collapsing it to X or Y or Zed.

We start with noisy seven, nine, or five
\\And end with perfect one. To better spot
\\Those flaws we must avoid, we first must strive
\\To find which ones commute and which do not.

With group and eigenstate, we've learned to fix
\\Your quantum errors with our quantum tricks.
\end{verse}

\end{document}